# Using Guilds to Foster Internal Startups in Large Organizations: A case study

Tor Sporsem[1][0000-0002-5230-7480], Anastasiia Tkalich[1][0000-0001-7391-4194], Nils Brede Moe[1][0000-0003-2669-0778], Marius Mikalsen[1][0000-0003-0882-7427] & Nina Rygh[2]

[1] SINTEF Digital, 7034 Trondheim, Norway
[2] DNV, 1361 Høvik, Norway

**Abstract.** Software product innovation in large organizations is fundamentally challenging because of restrained freedom and flexibility to conduct experiments. As a response, large agile companies form internal startups to initiate employ-driven innovation, inspired by Lean startup. This case study investigates how communities of practice support five internal startups in developing new software products within a large organization. We observed six communities of practice meetings, two workshops and conducted ten semi-structured interviews over the course of a year. Our findings show that a community of practice, called the Innovation guild, allowed internal startups to help each other by collectively solving problems, creating shared practices, and sharing knowledge. This study confirms that benefits documented in earlier research into CoPs also hold true in the context of software product innovation in large organizations. Henceforth, we suggest that similar innovation guilds, as described in this paper, can support large companies in the innovation race for new software products.

## 1    Introduction

Software product innovation is challenging in large organizations because they often lack the freedom to experiment and have established routines that limit flexibility [4]. Therefore, they need to find strategies to foster innovation [10]. One way is to establish a parallel organizational structure – like an Innovation Guild – to support employees innovating. Parallel structures perform functions that the regular organization does not or is ill-suited to perform well [13]. Some examples of parallel structures include quality circles [8] and *Communities of Practices* (CoP) [18]. Although some studies indicate that such parallel structures can boost innovation [16], their role in *software product innovation* is not well-examined.

Large organizations are currently trying Lean startup approaches to give internal startups the freedom to create new software products and experiment with customers, much like a standalone startup [4]. Innovation frameworks such as *The corporate startup* [17], design sprints in Google [21], "FedEx Day" and "20% Time" in Atlassian [10] are increasingly gaining attention as a way of giving guidelines and standardizing innovation processes to help organizations track and support software product innovation. Simultaneously, frameworks like these do state which practices or tools internal startups should leverage to drive innovation. However, internal startups are left to explore such practices and dig up needed knowledge themselves.

To shed light on this topic and recognizing that innovation in large agile companies may be particularly challenging, we ask the following research question: *How does a large organization use CoP to support internal startups in software product innovation?* To answer, we report on a case study of software product innovation at DNV Maritime, where a CoP – based on what Spotify call "Guilds" – was successfully applied to facilitate innovation processes inspired by Lean startup.

## 2   Related work

Parallel organizational structures, such as Communities of Practice (CoP), are commonly applied within software-intensive companies to support employees as problem-solving knowledge workers [13]. A community of practice is a *group of people who share a concern, a set of problems, or a passion about a topic, and who deepen their knowledge and expertise in this area by interacting on an ongoing basis* [18]. CoPs can take on different functions within an organization and evolve over time [11]. To an organization, a CoP can provide an arena for problem solving, drive strategic work, share best practice, onboard newcomers, develop professional skills, and start new lines of business [19]. To individuals, it can provide help in overcoming challenges, enabling contributing to your, improve professional reputation, provid a professional identity, and (most essential in our opinion) having fun.

In our case, DNV experimented with a CoP to support their employee-driven innovation. Without sufficient support, employee-driven innovation will fail [1]. Employees constantly need to improve their skills, share knowledge, and coordinate across the organization if they are to succeed.

Empirical studies of CoPs in software engineering are far between. Paasivaara and Lassenius [11] summarized some; [6, 7, 9]. Recently, research on the use of CoPs in Spotify (known as "guilds") have emerged [14]. This study identified four archetypes of CoPs:

- *Book clubs* focus on "learning instead of doing", where better working methods are discussed, but decisions are rarely made.
- *Open source societies* focus on members-owned components, maintaining them, improving, and finding strategies for them.
- *Support lines* focus on onboarding, providing answers to technical issues, and facilitating solutions discussions. Core experts guide less-experienced employees.
- *Standardizing committees* align practices across the organization by creating artifacts like toolset recommendations and standards.

Communities of practice are well-researched parallel structures. However, Paasivaara and Lassenius [11] argue that researchers need to study CoPs in new contexts to understand the concept further. In this study, we answer this call and examine CoPs in the context of software product innovation.

To understand how CoPs can support organizations as a parallel structure in DNV's case, we need to present some additional literature on innovation. *Software product innovation* is defined as the creation and introduction of novel software products to the market.

*Lean startup* is a popular approach to software product innovation where software is developed and validated through continuous experiments with customers to minimize development costs and increase customer satisfaction [12]. It is argued that the application of the Lean startup principles (e.g. *Build-Measure-Learn* and *validated learning*) increases the speed of product development [5] and improves product-market fit [3], but also faces challenges in large organizations [15].

So, how does a large company make use of Lean startup approaches? Large organizations foster *internal startups* [4] by encouraging new corporate efforts in their own environment to enter new markets and explore new business strategies [3]. One suggested solution is *The corporate startup* [17] which offers guidelines for software product innovation in an existing organizational environment. Innovation frameworks like Lean startup and The corporate startup are based on employee participation. People pitch their ideas, and the ones with the highest potential are prioritized.

## 3   Case description and research approach

Our case is the Maritime division of DNV, a large worldwide provider of business-to-business classification, certification, verification, risk management, training, and technical advisory services. DNV sets standards for ships and offshore structures that vessels in international waters must comply with, known as Class Rules. These rules comprise safety, reliability, and environmental requirements. DNV is operating globally and considers software products crucial for offering value to its worldwide customers. Hence, software product innovation has been part of the company's strategy to shift towards digital products and services. With 3 700 employees and headquarters in Hamburg, DNV Maritime has been using agile methods to develop software since 2008.

In 2018, the company established an innovation program based on the stage-gate innovation framework named The corporate startup [17]. Employees were invited to pitch ideas for new software products and created internal startups to develop them. These internal startups participated in a CoP, called the Innovation Guild, to support their innovative work, which is the focal point of this study.

We chose a case study [20] because we closely followed five internal startups in the between June 2020 and March 2021. We collected data in 3 different ways. First, we conducted seven interviews, asking internal startups how they work and their attitude towards guild meetings (two of them were interviewed twice). Then we did three interviews with managers on how they support the internal startups. Interviews were recorded and transcribed into 61 pages of text. Second, we collected observations from guild meetings and workshops by recording them and taking notes. Third, we used documentation on the innovation framework, such as strategic documents, status reports, and emails. Table 1 summarize our gathered data

**Table 1.** Data sources

| Data source | Description |
|---|---|
| Interviews | 10 semi-structured interviews (7 with internal startups, 3 with managers) |
| Meeting notes and transcripts | Guild (CoP) meetings (6 meetings, 90 minutes each), workshops (2 workshops, 2 days), venture board meetings (4 meetings) |
| Documents | Internal documents on organization, strategies and documentation of innovation framework implementation |

Data analysis was performed in three steps. First, textual data was entered into the qualitative data analysis tool NVivo. Two researchers coded the data inductively, which means that phenomenon and concepts rise from the textual data and make up themes/categories. Subsequently, we compared our categories with existing literature. We constructed codes separately followed by a comparison and discussion, ending up with a total of 150 codes. One example of a code: "Guild meetings helped me establishing contact to others with competence I needed." Further, we arranged the codes into 31 themes, e.g., "Cross functional cooperation contribute positively to internal startups" (which include the example-code above). As a last quality check, we presented our findings back to the informants. Comments were duly noted and cleared up small misconceptions.

The themes were grouped according to their impact on software product innovation. Which issue they addressed and how they supported internal startups is presented in table 3.

## 4     Results

DNV created and launched new products through the Innovation framework mentioned previously to facilitate software product innovation. The framework was based on a stage-gate model described in The corporate startup [17] and guided internal startups through six stages (Table 2) from ideation (*Customer insight*) to maturity (*Sustain*). Each product idea was suggested by an employee who became an *idea owner* and responsible for their own internal startup. They had to fulfill gate criteria to proceed from one stage to another (e.g., present evidence of the customer problem or customer intent). A group of business and domain experts (*Venture board*) evaluated whether the idea owners' evidence was sufficient to fulfill the criteria and progression. Operational line managers decided what amount of worktime idea owners could take out of their original job to work on the internal startups, varying from week to week – usually between 20-100 %.

Being originally operative specialists, the idea owners were unexperienced in entrepreneurship. It soon became evident that all internal startups faced common challenges and could draw on each others' knowledge to overcome them. Together with the innovation program manager, they decided to form a CoP – called *Innovation guild* – to share knowledge and find common solutions to the startups' shared problems. Besides, there was a need to establish connections to domain experts in other departments whom the internal startups had to rely on to develop their products.

**Table 2.** Stages of the innovation framework, criteria to proceed to next stage and key activities

| # | Stage | Criteria | Key activities |
|---|---|---|---|
| 1 | Customer insight | Evidence of the customer problem | Conduct customer interviews |
| 2 | Viability | Evidence of the customer intent | Build and test simple prototypes |
| 3 | Proof of concept | Evidence of feasibility for building, hard evidence of the customer intent | Build and test prototypes |
| 4 | Build | Evidence of possibility for scaling | Build MVP |
| 5 | Scale | Evidence for favorable market conditions | Marketing and sales campaigns, resource planning |
| 6 | Sustain | | Product improvement/sustain/retire |

A mandate was made in collaboration between idea owners and managers to justify the guild's existence: "sharing experiences, solving challenges, increasing competencies, providing access to expertise and finding new ways to interact with customers." Membership was primarily open for all internal startups. In addition, line managers and stakeholders from other units were invited to participate in guild meetings (depending on the topic of interest).

The idea owners chose topics based on shared challenges they were facing at the time and what they perceived valuable to discuss together. The guild gathered biweekly, with meetings approximately 1,5 hours long. Usually, the first 30 minutes were dedicated to idea owners sharing experiences since the last meeting, followed by the topic of interest (often presented by an invited external expert) before discussing what practices and knowledge were needed to drive innovation forward. A guild facilitator was in charge of planning the agenda and invited participants as the idea owners were far too busy handling their internal startups while juggling their departmental duties. Some weeks they worked full time on the startup, while some almost none, depending on how much their origin department allowed them.

The following subsections describe three distinct challenges that the Innovation guild was essential in solving (summarized in table 3). They are structured as a timeline, following the sequence of real-life events.

**Table 3.** Software product innovation challenges and achievements of the Innovation guild

| Challenge | Achievements of the Guild | Impact |
|---|---|---|
| Idea owners lacked customer contacts and know-how to approach customers | Acquiring common practices to approach customers in exploring customer-problems | Higher quality on feedback from customers and reduced time acquiring them |
| Lack of guidelines on pricing digital products, need to map the existing financial expertise | Increasing expertise in pricing digital products | Obtaining a pricing solution in line with the organization's existing strategy in less time |
| Insufficient knowledge on building and scaling products | Improving coordination with software development unit | Managers committed to dedicating developer resources earlier |

### 4.1 Acquiring common practices to approach customers

The first activity encouraged by the innovation framework was customer interviews (table 2, stage 1). However, customer insight was not an established practice among the idea owners. There was no systematic way of choosing or approaching the customers, and some did not even know who to contact. One idea owner commented: "It is the gut feeling that decides which company to approach. But how do we get a systematic way to get an insight on whom to approach?" In a Guild meeting, it was deciced to involve the marketing team to find a way, and in the following marketing and sales intelligence were invited to discuss the challenge. Three participants from the marketing team presented an overview of the tools they applied to communicate with customers and analyze markets (e.g., digital marketing and sales intelligence tool, customer segment, email templates). The idea owners found the meeting helpful; one of them commented: "For me the meeting was good. The customer matrix will help me to tune my email campaign." In this way, the guild assisted idea owners in acquiring new practices to approach customers by leveraging the marketing experts' existing knowledge and discussing ways of using it in the startups. As a result, they were enabled to achieve higher quality customer feedback in a reduced amount of time.

### 4.2 Building competence in pricing digital products

According to the innovation framework, idea owners had to present evidence of customers' intent to buy the new products (table 2, stages 2 and 3). How exactly such evidence could be collected was nonetheless unclear. One proposal was to demonstrate the customers' intent by collecting their feedback on tentative pricing models. However, idea owners held no expertise in pricing. The guild initiated a series of meetings inviting representatives from finance, digital sales, and line managers to address this challenge. Finance managers realized the need to identify what possibilities the existing payment mechanisms offered concerning the new digital pricing. He expressed: "Idea owners should suggest an idea on how their products can be priced, but it is important for us to find out which pricing models we can offer for them to choose from." In collaboration with digital sales experts, the finance managers created a list of available pricing models with instructions on how they fit different types of offerings. One idea owner explained: "A list of what pricing models are possible and not, is great. I am really happy to see that it is happening." In sum, the Innovation guild supported the internal startups by finding ways of pricing digital products through acting as a collaboration arena for idea owners and pricing experts. Henceforth, they saved time obtaining a pricing solution in line with the organization's existing pricing strategy.

### 4.3 Finding ways to collaborate with software developers

Entering the Build-stage (table 2, stage 4), the prototype of an internal startup was handed to the software unit for subsequent development. However, until this stage, the idea owners had focused only on exploring the business potential. Further, most idea owners did not hold sufficient knowledge on building and scaling products. They lacked documents that software developers needed to start developing, like feature lists and user stories. One idea owner stated: "You expect to give your idea to the IT guys and then come back in two weeks or a month, and everything is ready. But this is not how it turns out to be." After discussions in the guild, idea owners decided to include and coordinate with the software unit earlier in the innovation process. Software developers were invited to the following guild-meeting where they described their agile practices of working with new products in other business areas of DNV, followed by a discussion among idea owners on how they can fit this way of working into their innovation process. A second guild meeting was held on the topic where one idea owner and the head of the software department had successfully collaborated. As a result, other idea owners acquired knowledge of the software development process and found inspiration on how to make this collaboration work.

The line management also acknowledged that earlier involvement of software developers should be practiced whenever possible. They committed to dedicate software developers earlier. A line manager said: "We must avoid handover to IT and build as one team."

To summarize, the Innovation guild allowed internal startups to standardize practices to tackle shared challenges, build and share competence together, and create collaboration practices with other units. According to themselves, the Innovation guild helped internal startups to understand how to best practice innovation within DNV.

## 5  Discussion and conclusions

To innovate like startups, large agile companies need to develop strategies to foster software product innovation internally [10]. However, the application of ready-available guidelines are not sufficient alone to drive internal startups to excellence. One possible solution is to engage in Communities of practice [16] or other parallel structures that improve organizational problem-solving [11, 13]. To answer our research question – *How does a large organization use CoP to support internal startups in software product innovation?* – we described how a large agile company applied a CoP to foster internal starups. In the following discussion we summarize how the CoP succeeded and compare it to the types of guilds found in Spotify

Although innovation frameworks like The corporate startup [17] give step-by-step guidance on going from idea to product, we found that a parallel structure was needed to support it. Employees that usually carry out specialized tasks are an excellent source for new ideas. However, they typically lack experience in innovating software products. Idea owners needed to work together and draw on expertise from each other to solve problems that arose during developing collectively. The Innovation guild evolved into an arena where new practices emerged through sharing knowledge (e.g., how to develop and scale products), expand idea owners' skills (e.g., designing pricing models), improving coordination (with the software unit), and standardizing practice (such as how to work with the customers). In this way, CoPs can be seen as a prerequisite for succeeding with employee driven software product innovation that employees.

Our study's innovation guild holds similarities to communities of practice described by Wenger et al. [18], Paasivaara and Lassenius [11], and Smite et al. [14], who found that guilds support knowledge sharing, networking, and standardization. According to the archetypes identified in Spotify [14] (described in chapter 2), the innovation guild can be labeled as a *standardizing committee*. It helped idea owners create common practices and ways forward. Also, starting every guild meeting, idea owners shared their latest experiences, and guests were invited to share knowledge and inspire. This is typical for a *book club*. A mix between a book club and standardizing committee provides the startups with the best of two worlds: they can control what practices to standardize and acquire knowledge without committing to any decisions. From this, we can learn what type of guilds functions as a lubricant for innovation frameworks to work in large organizations, thus becoming a powerful organizational structure when innovating software products. Our study confirms that benefits documented in earlier research into CoPs [11, 16, 18] also hold true in the context of software product innovation in large organizations.

Despite large organizations struggling in reaching the same level of success as startups [4], they hold massive assets in terms of expertise, resources, and an established customer base. Innovation guilds activate these assets to support internal startups, hence contributing to their innovation process. In contrast, standalone startups have to establish inter-organizational alliances to access similar assets while risking their partners' opportunistic exploitation [2]. Thus, whereas startups ultimately face challenges alone, large organizations can become powerful allies for their internal startups when supported by parallel structures like an innovation guild.

Our study holds implications for innovation frameworks – and those using them – by describing how CoPs, like the Innovation guild, can supplement such frameworks in supporting internal startups. Simply implementing an innovation framework was insufficient in a large agile organization, while combining it with a parallel organizational structure like an Innovation guild drove innovation capability.